\documentclass[apjl]{emulateapj}
\usepackage{apjfonts}

\newcommand{\thi}{$\theta_{\rm i}$}
\newcommand{\teff}{\mbox{$T_{\rm eff}$}}

\accepted{July 19, 2005}

\begin{document}

\title{A New Technique for Probing Convection in Pulsating White Dwarf Stars}
\shorttitle{Seismic Probes of Convection}

\author{M. H. Montgomery}
\affil{Department of Astronomy, University of Texas, Austin, TX 78712, USA}
\email{mikemon@astro.as.utexas.edu}
\shortauthors{M. H. Montgomery}

\begin{abstract}
  
  In this paper we demonstrate how pulsating white dwarfs can be used
  as an astrophysical laboratory for \emph{empirically} constraining
  convection in these stars. We do this using a technique for fitting
  observed non-sinusoidal light curves, which allows us to recover the
  thermal response timescale of the convection zone (its ``depth'') as
  well as how this timescale changes as a function of effective
  temperature. We also obtain constraints on mode identifications for
  the pulsation modes, allowing us to use asteroseismology to study
  the interior structure of these stars.  Aspects of this approach may
  have relevance for other classes of pulsators, including the
  Cepheids and RR~Lyrae stars.

\end{abstract}


\section{Astrophysical Context}

The physics of convection represents one of the largest sources of
uncertainty in modeling stars. In main sequence objects, convection is
believed to occur in the cores of stars more massive than the Sun
\citep[e.g.,][]{Woo02} as well as in the envelopes of stars having
masses less than about $2.0 \,M_{\odot}$. Red Giant stars should have
fully convective envelopes \citep[e.g.,][]{Salaris02}, making
convection common throughout the H-R diagram. Along the white dwarf
cooling track we expect white dwarfs with helium spectra (DBs) and
temperatures less than $\sim\,$35,000~K to have surface convection
zones, while those with hydrogen spectra (DAs) and temperatures less
than $\sim\,15,000$~K should also have convective surface layers.

The fact that there are major uncertainties in our ability to model
the physics of convection has significant astrophysical consequences.
For instance, whether or not convective overshoot occurs in the cores
of massive stars affects the amount of material which is available for
nuclear burning, leading to an uncertainty of $\sim\,20\,$\% in
stellar ages \citep{DiMauro03,Bitzaraki01}. For pulsating white
dwarfs, uncertainty regarding convection in their atmospheres is the
largest single source of error in their derived effective temperatures
\citep[e.g., ][]{Bergeron95}. This is significant since we use what we
learn about the interior structure of the pulsators to calibrate white
dwarf cooling sequences, which in turn can be used to determine the
ages of individual white dwarfs \citep{Ruiz01} or the age of the
Galactic disc \citep{Wood98,Wood92,Winget87}.

The very low-amplitude oscillations which have been observed in the
Sun and recently in other Solar-like stars \citep[e.g., ][]{Bedding03}
have traditionally been explained through stochastic driving due to
the outer convection zones found in these stars
\citep[e.g.,][]{Kumar88,Houdek99}.  However, one of the first results
from the Canadian space mission \textsc{MOST} (Microvariability and
Oscillations of STars) has been that the oscillations expected in the
star Procyon are not present, at least not at detectable levels
\citep{Matthews04}, implying that our understanding of convection in
stars even slightly more massive than the Sun may still be incomplete.

In the sections which follow we describe a new technique for fitting
observed non-sinusoidal light curves in white dwarfs. With this
technique we can recover the thermal response timescale of the
convection zone (or its depth) and how this timescale changes as a
function of effective temperature. We also obtain mode identifications
for the pulsation modes, which helps us to use asteroseismology to
study the interior structure of these stars.

Our approach for deriving information on the depth of the convection
zone and its temperature sensitivity is based on the seminal numerical
work of \citet{Brickhill92a} and on the complementary analytical
treatment of \citet{Goldreich99a} and \citet{Wu01}. Essentially, we
take a hybrid of these two approaches, and it is this model which we
describe below.

\section{The Model}

\subsection{Assumptions}

\label{ass}

The fundamental assumptions of our hybrid model are that:
\begin{enumerate}
\item The flux perturbations beneath the convection zone are
  sinusoidal in time and have the angular dependence of a spherical
  harmonic.
\item The convection zone is so thin that we may locally ignore the
  angular variation of the nonradial pulsations, i.e., we treat the
  pulsations locally as if they were radial.
\item The convective turnover timescale is so short compared to the
  pulsation periods that the convection zone can be taken to respond
  ``instantaneously''.
\item Due to the extreme sensitivity of the convection zone to changes
  in temperature, we consider only flux and temperature variations,
  i.e., the large-scale fluid motions associated with the pulsations
  are ignored.
\end{enumerate}
The third assumption is given weight by appeal to standard convection
models of these stars, which indicate that the time taken for a
convective fluid element to circulate from top to bottom of the
convection zone should be less than a second. Since the pulsation
periods of these stars are much longer than this, i.e., hundreds of
seconds, this appears to be a safe assumption. While plausibility
arguments for the other assumptions may similarly be made, it is
perhaps better to see how well they allow us to model the observed
light curves before deciding on their viability (``the proof is in the
pudding'').

\subsection{Energetics}

We assume energy conservation in that the flux emitted at the
photosphere, $F_{\rm phot}$, is equal to the flux incident at the
base of the convection zone, $F_{\rm base}$, \emph{minus} the energy
per unit time which is absorbed by the convection zone,
$d\tilde{Q}/dt$, i.e.,
\begin{equation}
F_{\rm phot} = F_{\rm base} - \frac{d \tilde{Q}}{d t}.
\label{engy0}
\end{equation}
Physically, this can be thought of as the convection zone having a
specific heat; it must absorb some amount of energy for its
photospheric temperature to be raised by a given amount.

Due to the assumption that the convection zone is instantaneously in
quasi-static equilibrium, we can compute the term $d\tilde{Q}/dt$
purely in terms of static envelope models having different values of
$T_{\rm eff}$ \citep[see][ eq.~4]{Wu01}. This energy absorption rate
depends only on the photospheric flux (or $T_{\rm eff}$), leading to
the following equation:
\begin{equation} 
F_{\rm phot} = F_{\rm base} + \tau_C \frac{d F_{\rm phot}}{dt},
\label{engy}
\end{equation}
where the new timescale $\tau_C \equiv \tau_C \mbox{\scriptsize
  $(F_{\rm phot})$}$ describes the changing heat capacity of the
convection zone as a function of the local photospheric flux. Thus, we
have reduced the problem to a first-order, ordinary differential
equation in time\footnote{In the previous approaches of
  \citet{Ising01} and \citet{Brickhill92a}, one must essentially solve
  a second-order partial differential equation in time \emph{and}
  space, which is computationally much more intensive.}. Since the
perturbations at the base of the convection zone are assumed to be
linear, $F_{\rm base}$ is taken to have the usual time and angular
dependence of a nonradial oscillation mode, i.e.,
\begin{equation}
F_{\rm base} = {\rm Re}\,[\,A \,e^{i (\omega t + \delta)} Y_{\ell m}(\theta,\phi)],
\label{ansatz}
\end{equation}
so equation~\ref{engy} must be solved on a grid of points across the
visible surface of the star having different values of $\theta$ and
$\phi$, and the resulting fluxes need to be added together to
calculate the observed light curve (note: $A$ and $\delta$ in
equation~\ref{ansatz} are the amplitude and phase associated with the
pulsation mode, respectively).  Fortunately, because of the
simplifications we have introduced, this problem is still
computationally tractable.

\begin{figure}
  \centerline{\includegraphics[height=1.00\columnwidth,angle=-90]{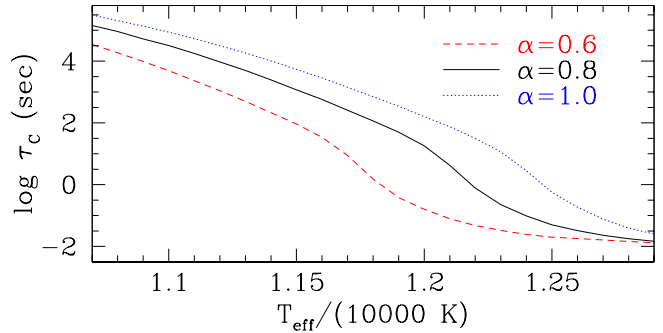}}
  \caption{Theoretical calculations of the thermal response 
    timescale of the convection zone, $\tau_C$, as a function of
    $T_{\rm eff}$. The local mixing-length theory of \citet{Bohm71}
    has been used (ML2), for the given values of $\alpha$.}
  \label{tauc}
\end{figure}

In Figure 1, we show a theoretical calculation of how this timescale
$\tau_C$ is expected to vary as a function of $T_{\rm eff}$. For a
given value of $\alpha$ (the mixing length to pressure scale height
ratio), we see that $\tau_C$ changes by approximately a factor of 1000
as the temperature goes from 12000~K (the observed onset of pulsations
in DAVs) to 11000~K (the observed disappearance of pulsations in
DAVs). If we parameterize this as a power law in $T_{\rm eff}$, i.e.,
\begin{equation}
  \tau_C \approx \tau_0 T_{\rm eff}^{-N},
  \label{tpar}
\end{equation}
we find that $N \approx$ 90--95 and $\tau_0$ is a function of
$\alpha$, demonstrating a \emph{very} strong temperature dependence
\citep[the curves in Fig.~\ref{tauc} were computed using the standard
mixing length theory of][]{Bohm71}. We note that although the value of
$\tau_0$ is affected by the version of mixing length theory employed
and the choice of $\alpha$, e.g., \citet{Bohm-Vitense58} versus
\citet{Bohm71}, the value of $N$ inferred from these standard models
is relatively constant, having values in the range 90--95 in the DAV
instability strip; for the DBVs, we find $N \approx$ 20--23.  The
simplified convection model used by \citet{Wu97}, while still quite
temperature sensitive, has a value of $N \approx 55$ for DAV models.
As we will see, the observations can be used to constrain the value of
$N$. For comparison with previous work by Wu, we note that in terms of
her variables that $N \approx 2\beta + \gamma$.

Since a typical pulsator may have excursions in temperature of several
hundreds of degrees, we expect $\tau_C$ to vary greatly during the
pulsations, leading to nonlinear light curves. Indeed, nonlinearities
beyond second order can become important, making it possible to
constrain not only $\tau_0$ but also the other parameters of the fit,
such as $N$, the inclination angle \thi, as well as the $\ell$ and $m$
values of the pulsation mode.

Since the fluxes in equations~(\ref{engy0})--(\ref{ansatz}) refer to
the total energy being gained and lost by the convection zone, they
represent bolometric fluxes.  By using sequences of static, numerically
computed envelopes to connect conditions at the base of the convection
zone with those at the photosphere, we implicitly have taken into
account the $T^4$ nonlinearities associated with variations in the
photospheric temperature. However, since the observations that are
typically made are in a finite wavelength range which is
well-separated from the peak wavelength flux, we must apply a
correction factor for the computed fluxes when comparing with the
observations.

For data taken with phototubes, we assume that the observations are
taken in a narrow bandpass centered on 4300~\AA, and for CCD data
taken with the ARGOS photometer we assume a central wavelength of
5000~\AA\ \citep{Nather04}. The simplified model for the correction
factor which we employ here assumes a blackbody distribution for the
radiative flux, which we then linearize about the effective
temperature; we apply this correction individually to each element of
surface area at each time step. In practice, this bolometric
correction leads to a reduction in the observed pulsation amplitudes
of $\sim 1.5$ for the DAVs and $\sim 2.4$ for the DBVs.  Finally, we
note that we have used the same simplified limb-darkening law as Wu
for both the DAVs and DBVs, namely the Eddington limb-darkening law:
\begin{equation}
  h(\mu) = 1 + \frac{3}{2} \,\mu,
\end{equation}
where $\mu$ is the cosine of the angle between a surface normal vector
and the line-of-sight.

\section{Lightcurve fitting}
\label{lfit}

In this section we examine how the hybrid model represented by
equations (\ref{engy})--(\ref{tpar}) compares with the model of
\citet{Wu97,Wu01}, both for synthetic data and for actual observations
of stars.

\begin{figure*}
  \centerline{
    \includegraphics[height=1.00\columnwidth,angle=0]{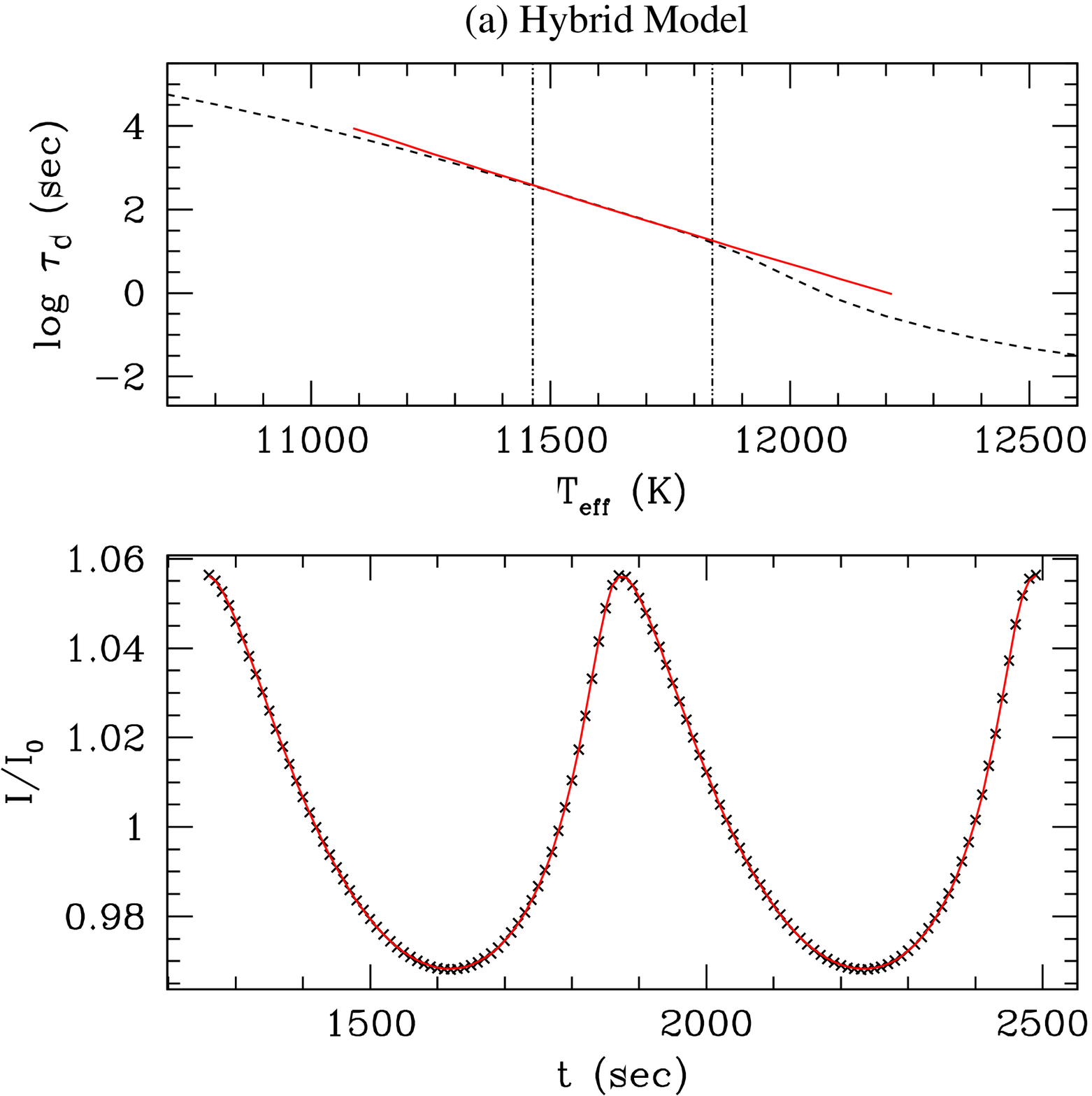}
    \hspace{0em}
    \includegraphics[height=1.00\columnwidth,angle=0]{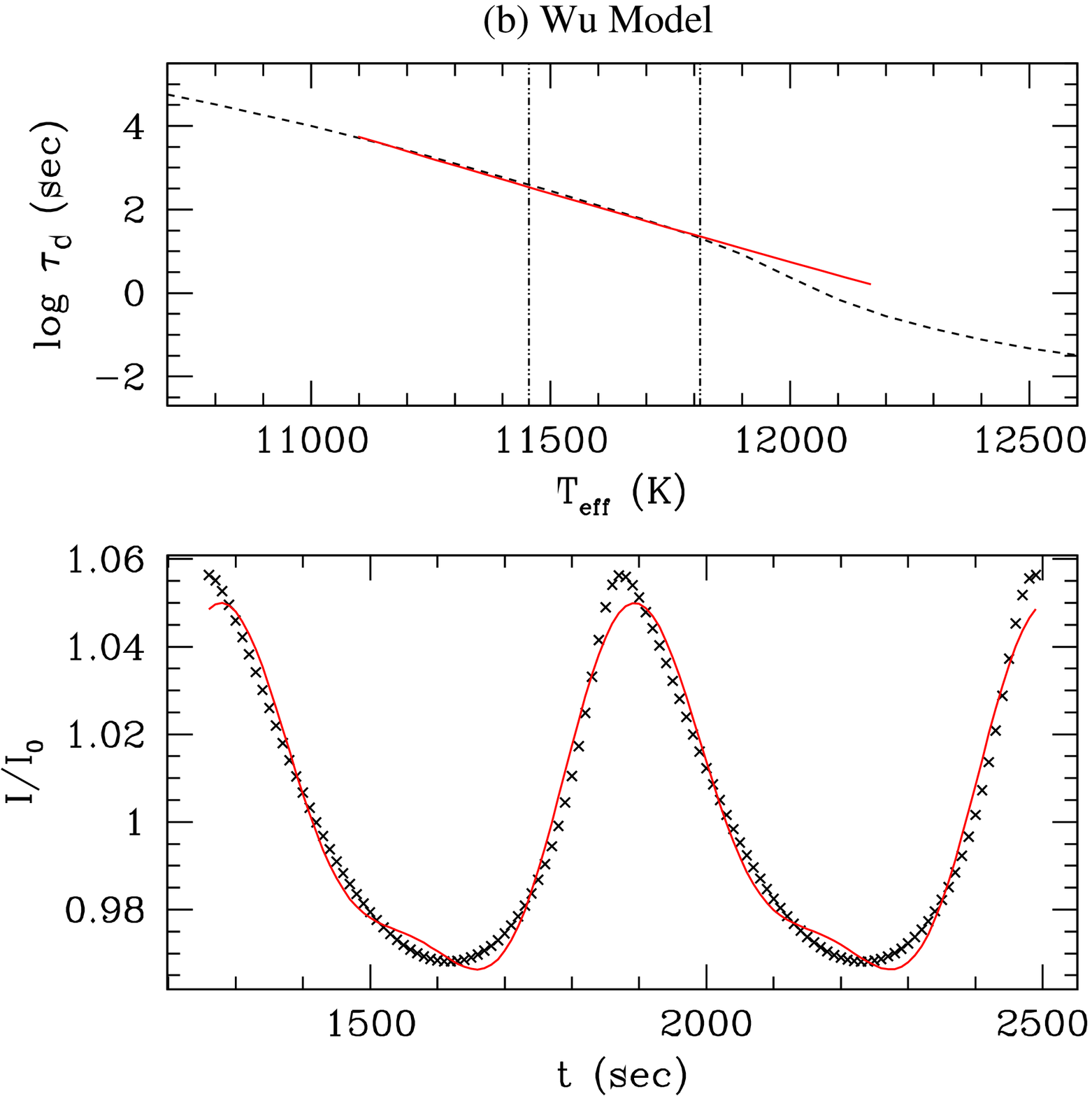}
  }
  \caption{Test fits of synthetic light curves having no angular dependence.
    In the lower panel of (a) we show the synthetic data (crosses) and
    our fit (solid curve), and in the upper panel we show the thermal
    response timescale used to generate the synthetic light curve
    (solid curve) compared with that which we derive from our fit
    (dashed curve). The vertical dotted lines indicate the maximum and
    minimum temperature excursions which the photosphere undergoes
    during a pulsation cycle. (b) The same as (a) but using Wu's
    second-order expression for the light curve fits.  
  }
  \label{testrad}
  \vspace*{-2em}
\end{figure*}

Wu's model uses the harmonics and combination frequencies that arise
from the second-order solution of equations
(\ref{engy})--(\ref{tpar}).  For a single mode with harmonics, Wu's
model constrains $\tau_0$, but the quantities $N$, \thi, and the
amplitude are degenerate with respect to each other. This is a result
of using only the lowest-order nonlinear terms.

Our approach is to numerically solve equations
(\ref{engy})--(\ref{tpar}), so we implicitly include the contributions
of all the nonlinear terms, including those of higher order. Thus, by
using more of the information present in the light curves we hope to
remove the degeneracies in the fits.

The technique developed by Wu was designed to be used for the
amplitudes and phases as derived from a Fourier transform of the data.
For consistency, we prefer to make comparisons directly with the light
curves, so we use her formulae to generate trial light curves which
are then fitted to the data.

\subsection{Tests omitting angular structure}

In this section we omit the angular structure and study fits to the
emergent flux from a fixed surface element using synthetic data; this
is formally equivalent to examining pulsations having $\ell=0=m$. In
addition to the convective parameters $\tau_0$ and $N$, we fit the
amplitude $A$ and phase $\delta$ in equation~\ref{ansatz}; the
frequency is assumed known from the observations. Since there is no
ambiguity associated with the choice of an inclination angle, we
expect that the method of Wu, along with our method, will lead to
non-degenerate best fit solutions.

In Fig.~\ref{testrad} we show fits to synthetic (test) light curves
using (a) our hybrid models and (b) Wu's second-order expression.  To
compute the synthetic light curve in this test we have assumed that
the thermal response timescale of the convection zone, $\tau_C$, is
given by local MLT models such as the curves shown in Fig.~\ref{tauc}.
Thus, the only difference between the input model and our fits is that
our fits assume the simplified form for $\tau_C$ given by
equation~\ref{tpar}. The adequacy of this assumption can be seen in
Fig.~\ref{testrad}a which shows that not only is the light curve
reproduced well (lower panel) but also that the match between the
input $\tau_C$ and that derived from the fit is quite good (upper
panel).  In Fig.~\ref{testrad}b, we show the same comparisons using
Wu's second-order expressions for the flux perturbations. In the lower
panel we see that this provides a reasonable qualitative fit to the
synthetic light curve, although the marked discrepancies between light
curve and the fit arise from neglecting the higher order terms.  In
the upper panel we see that the slope of the derived $\tau_C$ is
fairly close to that of the input model. 

\begin{deluxetable}{lcrrcccc}
  \tablecaption{Summary of test fits without noise
    \label{summary}}
  \tablehead{ 
    \colhead{model} & \colhead{\thi (deg)} &
    \colhead{$\tau_0$ (sec)} & \colhead{$N$} 
    & \colhead{Amp} &
    \colhead{$\ell$} & \colhead{$m$} & \colhead{MSD\tablenotemark{a}}}
  \startdata
  input model  &  ---  & 123.8 &  93.1  & 0.100 &  0   & 0  &   ---     \\
  hybrid model &  ---  & 122.5 &  94.9  & 0.100 &  0   & 0  &  0.014    \\
  Wu model     &  ---  & 109.0 &  89.7  & 0.088 &  0  &  0  &  8.272    \\[0.5em]
  \hline \\[-0.5em]

  input model      &  56.0 & 123.8 & 93.1   & 0.070 & 1 & 0    &   ---  \\
  hybrid model     &  54.5 & 122.3 & 99.5   & 0.067 & 1 & 0    &  0.001 \\
  hybrid model     & ~~0.0 & 118.4 & 71.4   & 0.067 & 2 & 0    &  0.021 \\
  Wu model         &  ---  & 111.2 & ---    & --- & --- & ---  &  1.343 \\
  hybrid model     &  83.1 & 162.0 & 359.4  & 0.094 & 1 & 1    &  ~~0.369 
  \enddata
\tablenotetext{a}{$
 \mbox{MSD} \equiv \frac{1}{N} 
      \sum_{i=1}^N \, [I_{\rm obs}(i)-I_{\rm calc}(i)]^2$ }
\end{deluxetable}

The parameters associated with these fits are presented in the upper
part of Table~\ref{summary}. The goodness of fit is measured by the
mean-squared deviation (MSD), defined by
\begin{equation} 
 \mbox{MSD} \equiv \frac{1}{N} 
      \sum_{i=1}^N \, [I_{\rm obs}(i)-I_{\rm calc}(i)]^2,
\end{equation} 
where $I_{\rm obs}$ and $I_{\rm calc}$ are the observed and
theoretically calculated light curves which have been normalized to an
average value of 1, respectively, and $N$ is the number of data
points. We see that the fit using the hybrid model is able to recover
the parameters of the input model quite well.  Although the Wu model
fit has \emph{much} higher residuals, it too reproduces the parameters
of the input model fairly well.

\begin{figure}
  \centerline{
    \includegraphics[height=1.00\columnwidth,angle=0]{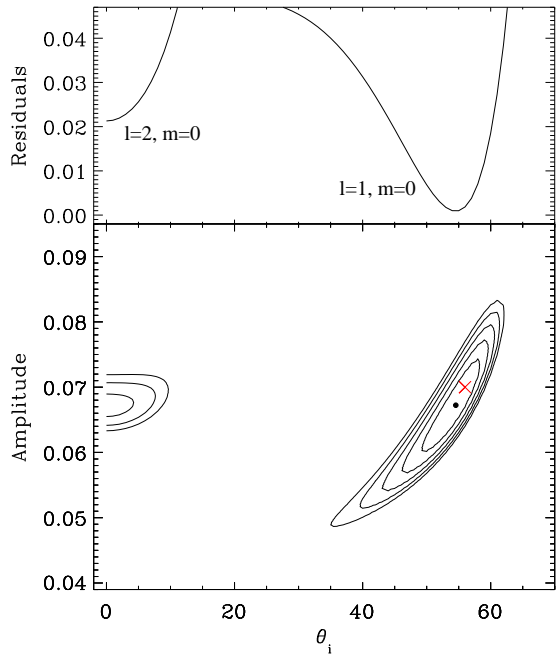}
    }
    \caption{Upper panel: The best fit residuals 
      as a function of \thi. Lower panel: Contours of constant fit
      residuals showing the relationship between the intrinsic
      amplitude of the mode and \thi. The minimum on the right
      corresponds to the $\ell=1$, $m=0$ solution, and the minimum on
      the left corresponds to the $\ell=2$, $m=0$ solution. The cross
      indicates the parameters of the input model.}
    \label{testtheta}
\vspace*{-1em}
\end{figure}

\subsection{Tests including angular structure}

\label{tang}

We now repeat the tests of the previous section, this time
incorporating the expected angular structure of the pulsation modes
into the calculation. In addition to the quantum numbers $\ell$ and
$m$ which now must be taken into account, the light curve shape also
becomes a function of the inclination angle, \thi. For the fits, we
have initially varied \thi\ in 1$^\circ$ increments, and in the
neighborhood of the minimum we have decreased the increments to
0.1$^\circ$.

In addition, we must discretize the surface of the stellar model and
sum over all elements to calculate the light curve. In order to
minimize the number of computations for a given accuracy, we
discretize the surface of the model so that the surface area of each
element projected along the line-of-sight is the same. Thus, elements
near the edge of the disc contribute with the same weight (except for
limb-darkening). Each of the elements is wedge-shaped and they are
arranged radially on annuli centered on the the visible disc of the
model. The input models were calculated using 1024 surface elements
arranged on 16 annuli, and the subsequent fits to these models and
fits elsewhere in this paper have used 256 elements arranged on 8
annuli.

In Wu's second-order theory, a degeneracy exists among the parameters
such that it is not possible to disentangle \thi\ from the intrinsic
mode amplitude or from the thermal timescale parameter $N$, i.e.,
equally good fits are possible for any value of \thi.  Our hope is
that the inclusion of the higher-order nonlinear effects in our hybrid
model will lift this degeneracy and allow unique fits to be made.

In the lower part of Table~\ref{summary} we show the results of
different fits to a sample light curve having $\ell=1$, $m=0$, and
\thi\ $=56.0^\circ$ (these values are similar to those derived in
Section~\ref{PG1351} for PG1351+049).  Based on the residuals, the
hybrid model with $\ell =1$, $m=1$ can be ruled out, as can be the Wu
model, although it again provides a reasonable estimate of $\tau_0$.
The remaining hybrid models both have very low residuals, making a
choice between the two difficult.

\begin{figure}
  \centerline{
    \includegraphics[height=1.00\columnwidth,angle=0]{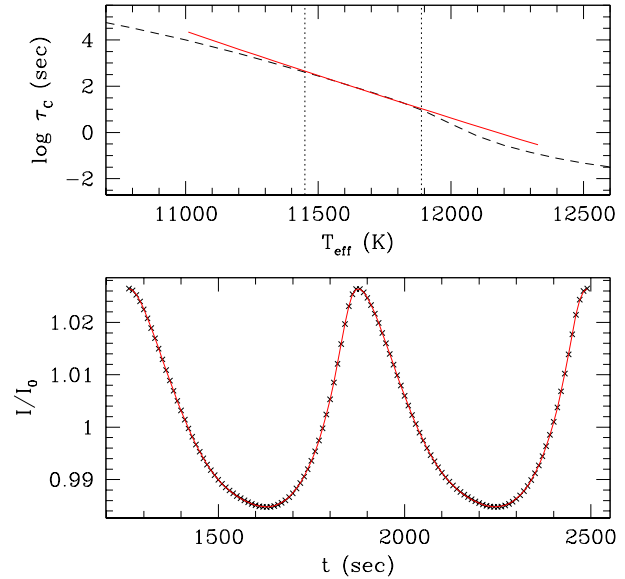}
  }
  \caption{The same plot as Fig.~\ref{testrad} but for the best fit solution
    of Figure~\ref{testtheta}.  }
  \label{testang}
\vspace*{-1em}
\end{figure}

In Figure~\ref{testtheta} we examine the uniqueness of the mode
identifications as well as possible degeneracies in the fit
parameters.  In the upper panel the residuals of the two best fits are
plotted as a function of \thi, and in the lower panel contours of
constant residuals of these fits are shown as a function of \thi\ and
mode amplitude.  We see that while there is a partial degeneracy
between \thi\ and the amplitude $A$, the presence of the higher-order
nonlinearities does indeed lift this degeneracy and allow a unique
solution to be obtained.  This best-fit solution is shown in
Figure~\ref{testang}.

Although no noise has been added to the synthetic light curve, the
derived value of \thi, 54.5$^\circ$, does differ slightly from the
input value of 56.0$^\circ$. This is because for finite amplitude
pulsations the parameterization in equation~(\ref{tpar}) for $\tau_C$
does not exactly reproduce the variations in $\tau_C$ of the input
model.

From the above fits, we see that a mode with $\ell=1$, $m=0$, and
$\mbox{\thi} = 56^\circ$ can be mimicked by a mode with $\ell=2$,
$m=0$, and \thi$=0^\circ$; indeed, even the amplitudes which are
derived in these two cases are nearly identical. While more work is
needed to address the cause of this similarity, we find from
exploratory calculations that the degeneracy between these mode pairs
exists only when $\mbox{\thi} > 45^\circ$ for the $\ell=1$ mode.

\begin{deluxetable*}{rccccccc}
  \tablecaption{Summary of test fits including noise
    \label{noise_sum}}
  \tablehead{ 
    \colhead{model} & \colhead{\thi (deg)} &
    \colhead{$\tau_0$ (sec)} & \colhead{$N$} 
    & \colhead{Amp} &
    \colhead{$\ell$} & \colhead{$m$} & \colhead{MSD}}
  \startdata
  input model  &  56.0 &           123.8 &             93.1   &         0.070 &             1 & 0 &   ---  \\
  hybrid model &  51.5 $\pm$ 5.2 & 127.1 $\pm$ ~~9.0 & -108.6 $\pm$  16.9 & 0.066 $\pm$ 0.009 & 1 & 0 & 0.459 $\pm$ 0.056 \\
  hybrid model &~~0.3  $\pm$ 0.5 & 118.5 $\pm$ ~~6.5 & ~~-69.8 $\pm$ ~~2.4 & 0.067 $\pm$ 0.002 & 2 & 0 & 0.437 $\pm$ 0.081 \\
  hybrid model &  84.0 $\pm$ 1.3 & 212.6 $\pm$  51.4 & -355.1 $\pm$  10.1 & 0.111 $\pm$ 0.019 & 1 & 1 & 1.711 $\pm$ 0.168 \\
  hybrid model &  85.1 $\pm$ 1.4 & 104.2 $\pm$  11.5 & ~~~~-3.0 $\pm$ ~~1.8 & 0.620 $\pm$ 0.148 & 2 & 1 & 1.779 $\pm$ 0.153 \\
  hybrid model &  0--90          &          100--110 &    0               & 0.16--0.40        & 2 & 2 &           10--11    
  \enddata
\end{deluxetable*}

\begin{figure}
  \centerline{
    \includegraphics[height=1.00\columnwidth,angle=0]{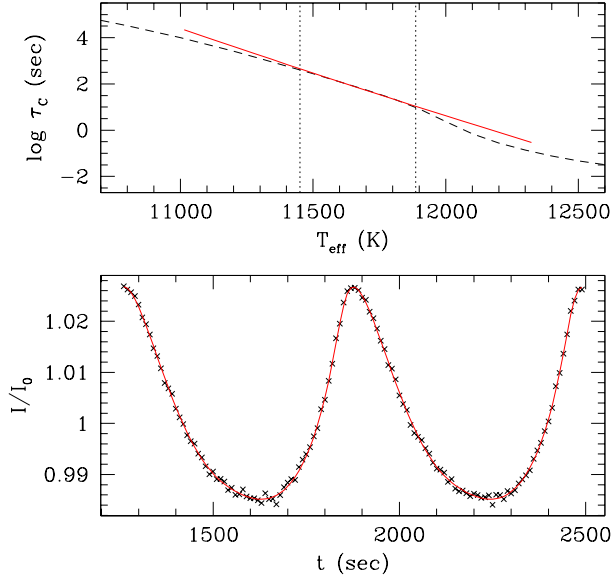}
  }
  \caption{The same plot as Fig.~\ref{testang} but for Gaussian noise
    added to the synthetic light curve. }
  \label{noise}
  \vspace*{-2em}
\end{figure}

\subsection{Tests including noise}

In this section we examine the effect of Gaussian noise on our light
curve fits. In Figure~\ref{noise} we show an example of such a fit,
where the level of the noise has been chosen to be somewhat higher
than the data which will be analyzed in section~\ref{fits}. In
Table~\ref{noise_sum} we show the results of many different fits to
the given input model for solutions having different $\ell$ and $m$
values. The parameter values and their error bars listed in the table
were obtained by computing the average and standard deviation of fit
values for 10 different realizations of the noise. In the case of the
$\ell=2$, $m=2$ fit, the fits are so poor that there is no single
``best fit''; for this case, we have simply indicated the allowed
ranges for the parameters.

Unsurprisingly, the two fits having the lowest residuals are the same
ones as in the noise free case, and the difference in their residuals
is not statistically significant. Thus, choosing between them on the
basis of their residuals alone is not possible, and we are indeed
faced with the problem of non-uniqueness. As in the noise free case,
however, these fits still provide a good estimate of $\tau_0$ and the
amplitude, and the $\ell=2$ solution gives a decent estimate of $N$.

Examining the results of this and the previous sections, we see that
the hybrid model is able to accurately derive many of the parameters
of the input model, both for the case with and without noise. Indeed,
the model of Wu can also provide reasonably good estimates of
$\tau_0$, although not of the other parameters due to the degeneracy
between \thi, $N$, and the amplitude.  While fits to actual data will
of course be more difficult, these results give us hope that we will
be able to derive meaningful constraints on the parameters through
such a fitting procedure.

\subsection{Fits to observations}

\label{fits}

Since we are using a new technique, we wish to start with the best
possible candidates, i.e., those for which the application of the
method is the most straightforward.  For the present application, we
therefore restrict ourselves to stars which (1) are
\emph{mono-periodic}, having one oscillation mode which dominates
their light curves, (2) have \emph{large-amplitudes}, so that their
light curves contain clear nonlinearities, and (3) have \emph{high
  signal-to-noise data}, so that the errors are as small as possible.
In future work we will examine the degree to which these criteria can
be relaxed, while the focus of our present work is to demonstrate that
the method does indeed work for the cases examined below.

While by no means an exhaustive list, two stars fill these
requirements admirably: the DBV PG1351+049 and the DAV G29-38. Both
stars have light curves which are nearly mono-periodic (at least during
the time the observations were made), have significant nonlinearities,
and high quality data are available for both. In the following two
sections we describe the fits to these data.

\begin{figure*}
  \centerline{
    \includegraphics[height=1.0\columnwidth,angle=-90]{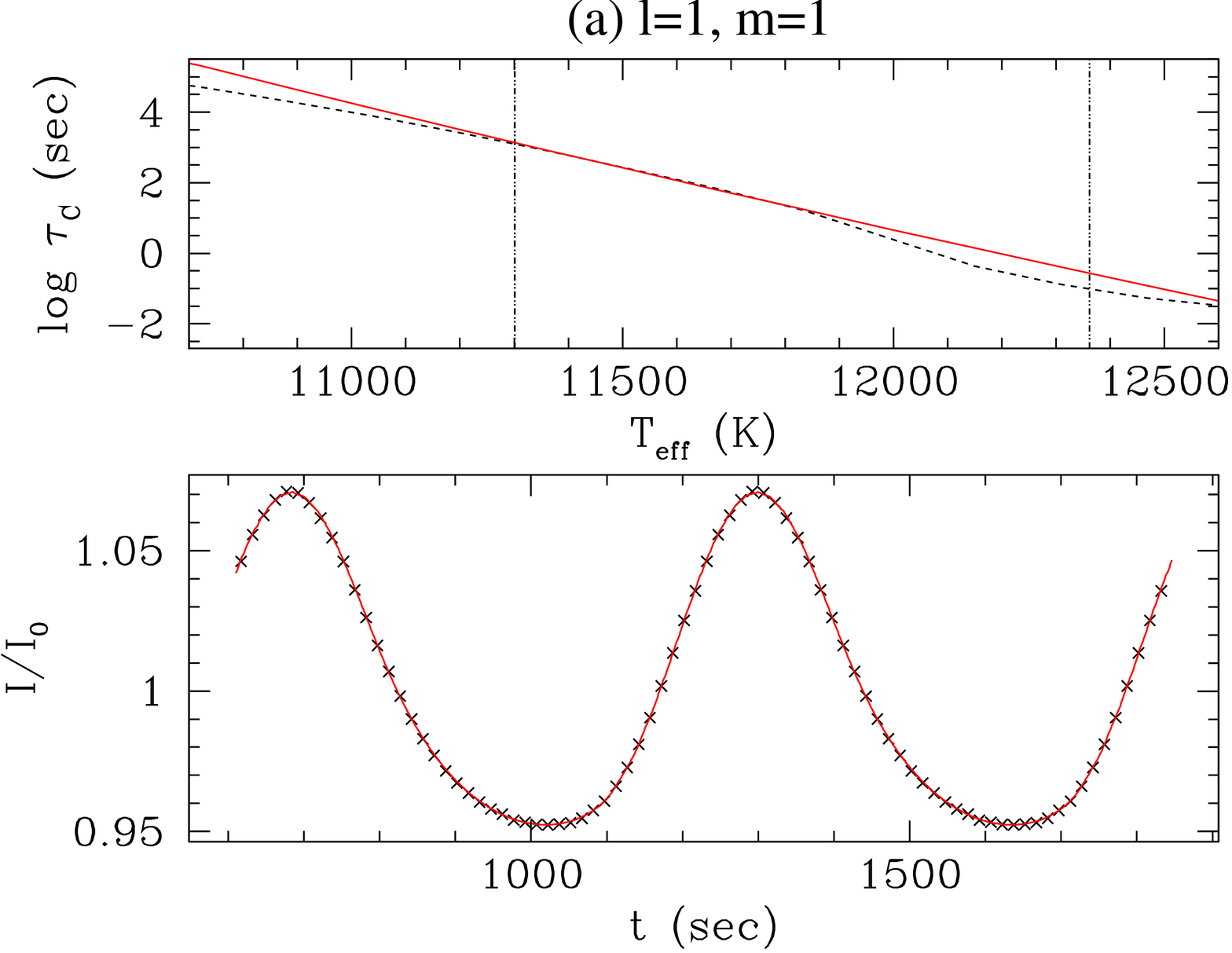}
    \hspace{0.0\columnwidth}
    \includegraphics[height=1.0\columnwidth,angle=-90]{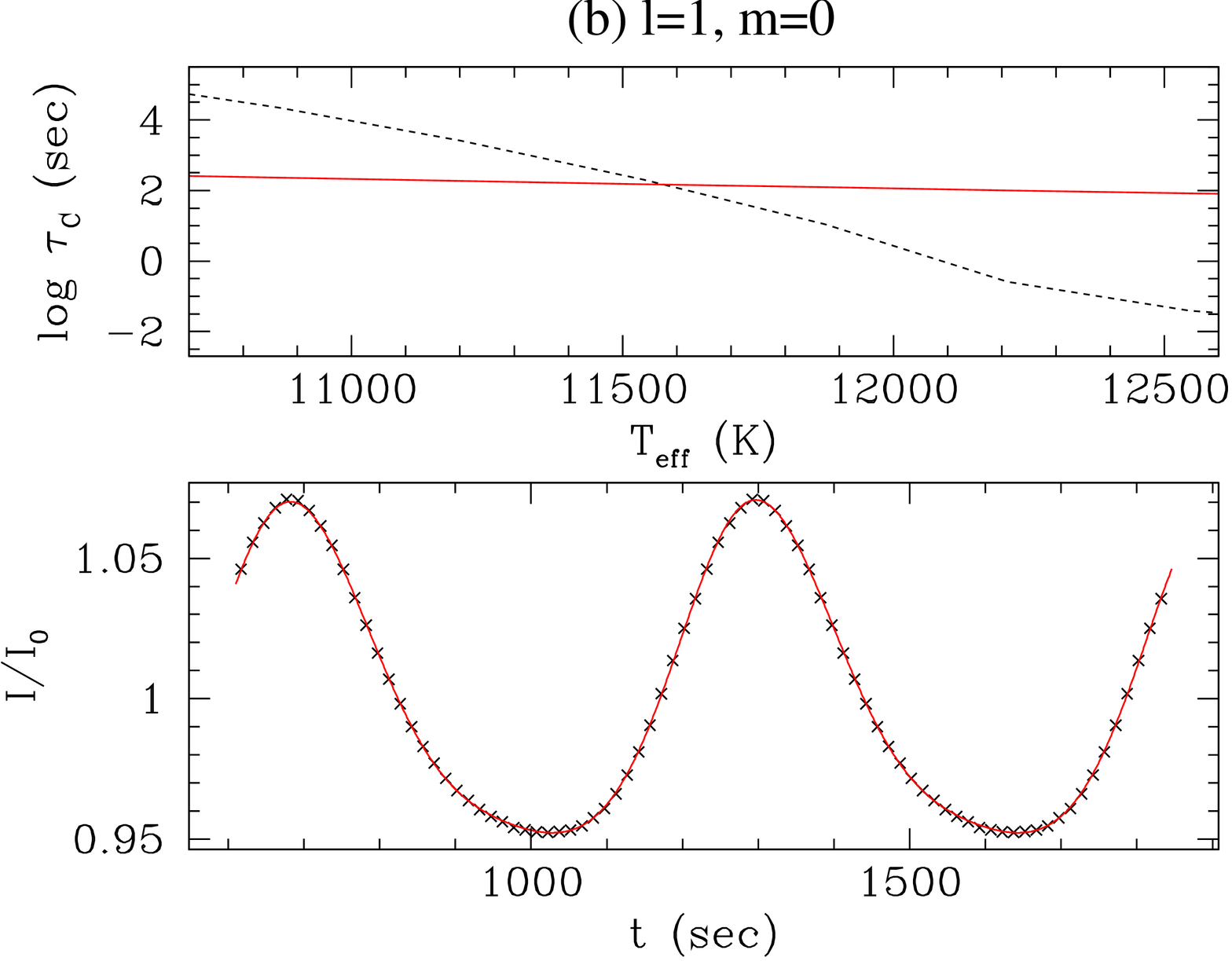}
 }
  \caption{Lower panels: fits (solid curve) to the observed pulse shape 
    (crosses) of the DAV star G29-38 for observations taken in 1988.
    Upper panels: the empirically derived $\tau_C$ as a function of
    $T_{\rm eff}$ (solid curve) versus that calculated assuming a
    given mixing-length theory (dashed curve). The fit on the left
    assumes $\ell=1$, $m=1$, while the fit on the right assumes $\ell=1$
    and $m=0$.}
  \label{g29fit}
\end{figure*}

\begin{deluxetable*}{ccccccc}
  \tablecaption{Derived parameters for G29-38
    \label{g29} }
  \tablehead{ 
    \colhead{\thi (deg)} &
    \colhead{$\tau_0$ (sec)} & \colhead{$N$} 
    & \colhead{Amp} &
    \colhead{$\ell$} & \colhead{$m$} & \colhead{MSD}}
  \startdata
  65.5\tablenotemark{a}$\!\!$ $\pm$ 3.4 & 187.4 $\pm$ 20.3 &  95.0 $\pm$ 7.7 & 0.259 $\pm$ 0.011 &  1 & 1  &  0.160 $\pm$ 0.045 \\
  73.9 \,$\pm$ 0.7 & 150.1 $\pm$ ~~6.5 & ~~7.1 $\pm$ 0.6 & 0.417 $\pm$ 0.025 &  1 & 0  &  0.178 $\pm$ 0.035 \\
   31.9 \,$\pm$ 1.3 &  151.2 $\pm$ ~~7.4 & ~~5.8 $\pm$ 0.6 & 0.363 $\pm$ 0.022 & 2 & 0 & 0.171 $\pm$ 0.037 \\
   82.0  \,$\pm$ 0.8 &    139.4 $\pm$ ~~3.9 &  ~~1.6 $\pm$ 0.4 & 1.190 $\pm$ 0.098 & 2 & 1 & 0.382 $\pm$ 0.028 \\
   $\sim$80.0  &  $\sim$110 & 0.0 & $\sim$0.27 & 2 & 2 & $\sim$41.0 
\enddata
\tablenotetext{a}{preferred fit}
\end{deluxetable*}

\subsubsection{The DAV G29-38}

The light curve of the DAV G29-38 was obtained by S. Kleinman in 1988
\citep{Winget90,Kleinman95}. Due to the very long time baseline of the
observations, the resulting folded light curve/pulse shape has a very
high signal-to-noise. We show this light curve, folded at a period of
615.15~s, along with two different fits to it, in the lower panels of
Figure~\ref{g29fit}: the crosses are the folded light curve and the
solid curve are the fits to it. The fit on the left is for an
$\ell=1$, $m=1$ pulsation mode, while that on the right is for an
$\ell=1$, $m=0$ mode.

From Table~\ref{g29} we see that these two fits, along with the
$\ell=2$, $m=0$ fit, have residuals which are not statistically
different from one another. However, they imply vastly different
temperature sensitivities (i.e., $N$) for the thermal response
timescale of the convection zone. In particular, if white dwarf
convection zones were as weak a function of \teff\ as implied by
either the $\ell=1$, $m=0$ or the $\ell=2$, $m=0$ solution ($N \sim$
5--7), there would hardly be any change in depth of their convection
zones as they cross the ZZ Ceti instability strip. This runs counter
to all theoretical predictions of convection, whether based on MLT or
on numerical hydrodynamic simulations \citep{Freytag95}. We conclude
that the physically meaningful solution is the $\ell=1$, $m=1$
solution. This yields a value of $N\approx 95$, which is exactly in
the range predicted by standard models of convection. In addition, we
note the identification $\ell=1$ is consistent with that found by
\citet{Clemens00} from an examination of the wavelength dependence of
the pulsation amplitude.

\subsubsection{The DBV PG1351+049}

\label{PG1351}

In 1995 the DBV star PG1351+049 was observed as part of a Whole Earth
Telescope \citep[WET; ][]{Nather90} campaign
\citep[e.g.,][]{Winget91,Winget94}. In 2004, we re-observed this star
using the Argos CCD photometer on the McDonald 2.1m telescope.  This
second set of high quality observations is important because it allows
us to cross check this technique. If we really are learning about the
fundamental parameters of the star then we would not expect those
parameters to have changed, i.e., the average depth of the convection
zone, $\tau_0$, and the inclination angle, \thi, should be the same,
although other quantities such as the amplitude of the mode might very
well be different.

As we saw in the previous section, a degeneracy in the parameters can
occur so that two sets of parameters produce roughly equivalent fits.
This is also the case for this star; we find both an $\ell=1$, $m=0$ fit
and an $\ell=2$, $m=0$ fit which reproduce the data well.

\begin{figure*}
\centerline{
  \includegraphics[height=1.0\columnwidth,angle=-90]{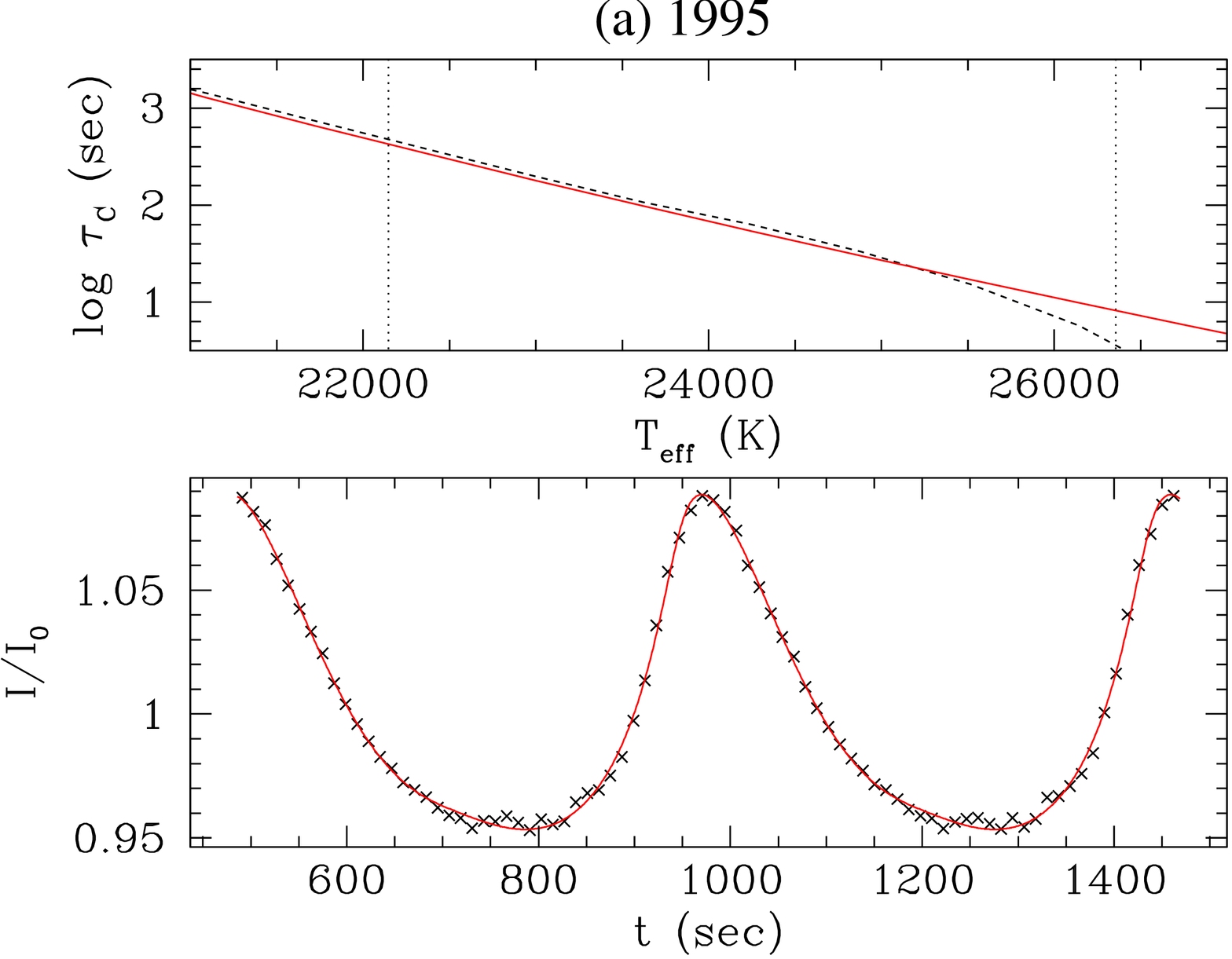}
  \hspace{0.00\columnwidth}
  \includegraphics[height=1.0\columnwidth,angle=-90]{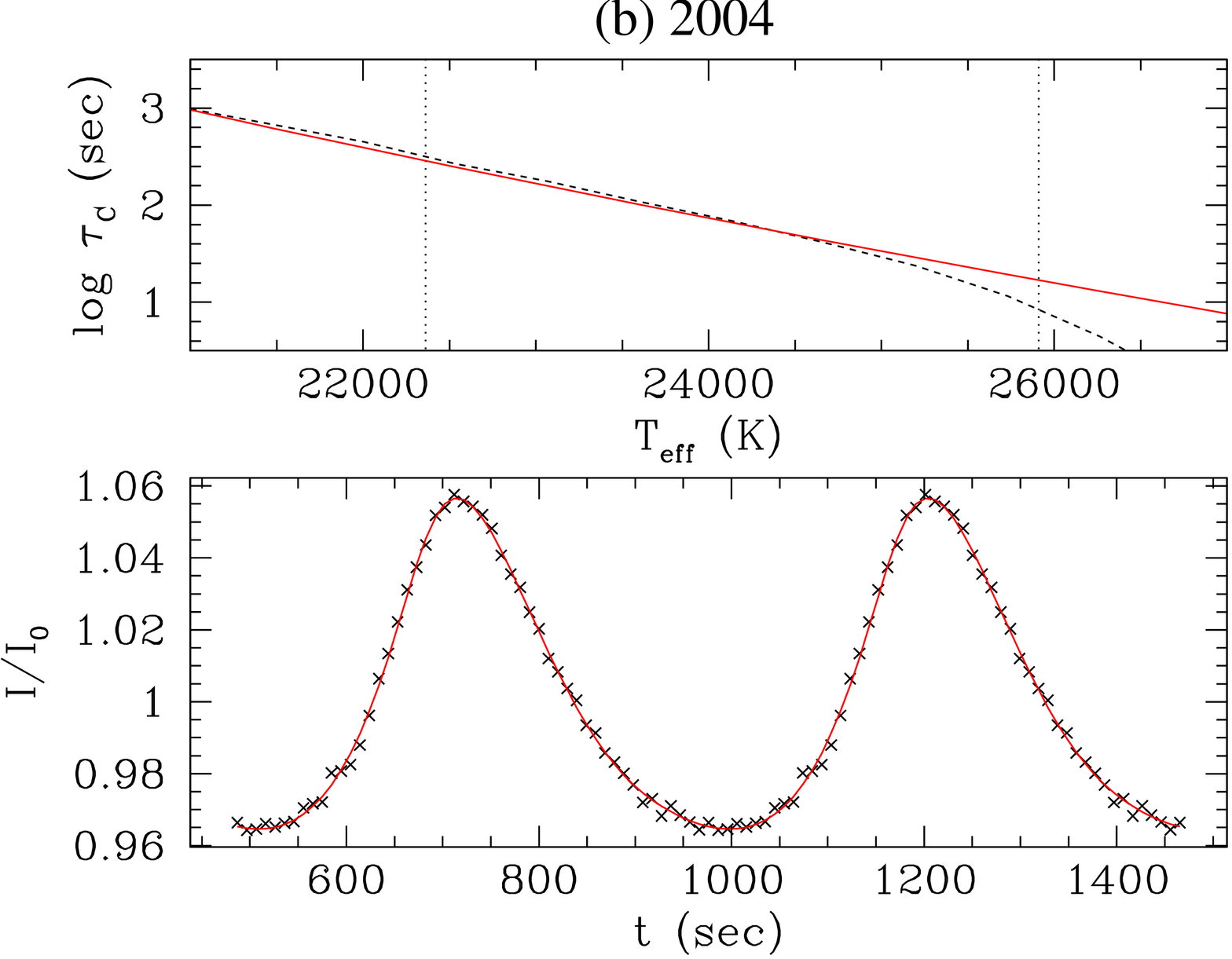}
 }
  \caption{Fits to the observed pulse shape 
    of the DBV star PG1351+049 for (a) data taken in 1995 with the
    Whole Earth Telescope (WET) and (b) data taken in May 2004 with
    the Argos photometer on the McDonald 2.1 m. Both fits assume
    $\ell=1$, $m=0$. We note that the star's pulsation amplitude and
    pulse shape was different in these two epochs.}
  \label{pg1351}
  \vspace*{0em}
\end{figure*}

In Figure~\ref{pg1351}, we show the results of the $\ell=1$ fits.  The
data have been folded at the period of the mode, 489.335~s, to improve
signal-to-noise and to simplify the fitting.  We note that the fit is
very good and reproduces the features in the light curve quite well.
In the upper panel we show the convective timescale $\tau_0$ derived
from these data (solid curve) and one computed using mixing-length
theory (dashed curve).  The fact that the slopes are similar is
encouraging, since the theoretical curve can be moved upward or
downward simply by tuning the value of $\alpha$, while its slope is
much less sensitive to $\alpha$.

From Table~\ref{pgcomp}, we see that the results of the $\ell=1$ fits
are remarkably consistent between the two sets of observations, with
the main difference being that the inferred amplitude has decreased by
approximately 30\%. In other words, even though the pulse shapes of
PG1351+049 were different in the two epochs, they can both be fit by
solutions whose only significant difference is the amplitude of the
mode. This is further evidence that the proposed mechanism, i.e.,
modulation of the flux by the convection zone
\citep[e.g.,][]{Brickhill92a,Wu01}, is the dominant nonlinear effect.

From Table~\ref{pgcomp}, we see that while the $\ell=2$ solution has
marginally lower residuals it is essentially equivalent to the
$\ell=1$ solution. Interestingly, we see that $\tau_0$, $N$, and even
the amplitude of the $\ell=2$ mode are reasonably close to the values
derived from the $\ell=1$ solution. The only large difference is the
value of the inclination angle, which is 0$^{\circ}$ for the $\ell=2$
fit as opposed to $\sim 58^{\circ}$ for $\ell=1$. While such a pole-on
viewing angle is certainly possible, it is much less likely than a
more equatorial orientation. For instance, given random orientations,
the probability of $\theta_{\rm i} < 10^{\circ}$ is only 1.5\%. For
this reason, the $\ell=1$ fit seems the likelier choice, although we
cannot rule out the $\ell=2$ possibility. Given that $\tau_0$ and $N$
are very similar in both fits, we are still able to obtain constraints
on the physics of convection using these data.

\begin{deluxetable*}{lrcrcccc}
  \tablecaption{Derived parameters for PG1351+049
    \label{pgcomp} }
  \tablehead{ 
    \colhead{epoch} & \colhead{\thi (deg)} &
    \colhead{$\tau_0$ (sec)} & \colhead{$N$} 
    & \colhead{Amp} &
    \colhead{$\ell$} & \colhead{$m$} & \colhead{MSD}}
  \startdata
  1995  &  57.8 $\pm$ 1.6 & 86.7 $\pm$ 8.3 &  22.7 $\pm$ 1.3 & 0.328 $\pm$ 0.018 &  1   & 0  &  4.15 $\pm$ 0.65   \\
  2004  &  58.9 $\pm$ 3.1 & 89.9 $\pm$ 3.6 &  19.2 $\pm$ 2.1 & 0.257 $\pm$ 0.021 &  1   & 0  &  0.95 $\pm$  0.25  \\
  1995  &   0.0 $\pm$ 5.9 & 85.1 $\pm$ 8.8 &  18.1 $\pm$ 1.4 & 0.305 $\pm$ 0.014 &  2   & 0  &  4.04 $\pm$ 0.74 \\
  2004  &   0.0 $\pm$ 6.1 & 89.0 $\pm$ 6.1 &  16.0 $\pm$ 1.1 & 0.233 $\pm$ 0.011 &  2   & 0  &  ~0.94  $\pm$ 0.15 
\enddata
  \vspace*{0em}
\end{deluxetable*}

\section{Discussion}

The fits we have obtained for the stars PG1351+049 and G29-38 are
quite good, implying that the physical assumptions which have gone
into our models are realistic. Although these fits are not necessarily
unique, for G29-38 we were able to rule out on general theoretical
grounds the one competing solution, and for PG1351+049 we found that
both solutions gave similar values of the parameters (other than the
inclination angle). Thus, if we view this as a technique to determine
the parameters $\tau_0$, $N$, and the mode amplitude then we have been
successful.

Previous studies using the approach of \citet{Wu01} to determine the
various parameters describing convection and pulsation
\citep[e.g.,][]{Kotak02} have met with less success than our present
analysis. We believe this is for three reasons. First, by using the
higher order nonlinearities we are using more information in the light
curve, which lifts some of the degeneracies. Second, the stars we
examined had large amplitudes (i.e., 5--10\%) which made the nonlinear
effects larger and easier to measure. Third, after being folded the
light curves of our objects had very high signal-to-noise.

We note that from the fits to PG1351+049 and the test cases examined
in Section~\ref{tang}, this method of mode identification provides
better constraints on $m$ than on $\ell$. This is fortunate, since the
complementary technique of chromatic amplitudes
\citep{Clemens00,Robinson95} provides constraints only on $\ell$.
Therefore, using these techniques in concert, it should be possible to
obtain unique mode identifications.

Our results also seem to make sense in terms of simple (and naive)
theories of convection, as can been seen by the close agreement
between the slopes of the dashed and solid curves in the top panels of
Figures~\ref{g29fit}a and \ref{pg1351}; thus, when combined with
spectroscopic temperature estimates they can lead to estimates of the
mixing length for a given convection model.  In addition, these fits
can provide independent constraints on the $\ell$ and $m$ values of
pulsation modes, something which is necessary for using
asteroseismology to study the interiors of these stars.  We consider
these results a proof of concept for this method, and they show that
pulsating white dwarfs may provide the ideal crucible for tests of
theories of convection.

In deriving these fits, we have made no assumptions about either the
temperature of the star or the value of any mixing-length parameter
which might be used to describe its convection zone.  If we now wish
to assume that we know the $T_{\rm eff}$ and $\log g$ values for these
stars, for instance, by taking values from the literature, then
we can derive further constraints on the value of $\alpha$ for a
particular mixing length theory. 

For the DAV G29-38, \citet{Bergeron04} derive the values $T_{\rm eff}
= 11,820$~K and $\log g=8.14$. Assuming ML2 convection we find that
this implies a value of $\alpha \sim 0.6$. For the DBV PG1351+049, if
we take $T_{\rm eff} = 22,600$~K and $\log g=7.9$ \citep{Beauchamp99},
then we find a value of $\alpha \sim 0.5$.  As a point of comparison,
\citet{Montgomery04a} found, albeit for hotter models, that the peak
convective flux in their non-local Reynolds Stress model could be
approximately reproduced for $\alpha \sim$ 0.6--0.7 for the DA models
and $\alpha \sim$ 0.4--0.5 for the DB models, which is broadly
consistent with the above values.

Since the value of $\alpha$ is a parameterization of the complex
problem of turbulent heat transport, there is no reason to suppose a
priori that the value of $\alpha$ needed to represent the physics in
the ``efficient'' regime near the base of the convection zone will
also be appropriate for representing the physics in the
``inefficient'', optically thin region near the photosphere. Thus, the
fact that our studies imply $\alpha \sim 0.5$ for the DBVs is not
necessarily at odds with the value of $\alpha = 1.25$ which is assumed
in the model atmosphere fits of \citet{Beauchamp99}. On the other
hand, the value $\alpha=0.6$ which we derive for the DAVs appears to
be consistent with that used for many model atmosphere fits of these
stars \citep{Bergeron04}.

\newpage

\section{Conclusions}

Encouraged by our successes with these two stars, we would like to
apply this method to pulsating white dwarfs throughout both the DAV
and DBV instability strips.  Thanks to the Sloan Digital Sky Survey,
the number of known white dwarf pulsators has essentially doubled in
the last two years \citep{Mukadam04a,Kleinman04,Mullally04}. Since
these pulsators, as well as those previously known, are a population
having a range of temperatures and masses, we will be able to map out
the depth of their convection zones as a function of both $T_{\rm
  eff}$ \emph{and} $\log g$. This should provide us with a detailed
map of how convection works in both the DAV and DBV instability
strips.

Given that most pulsating white dwarfs are multi-periodic, this
technique needs to be extended to deal with pulsators which have more
than one pulsation mode simultaneously present. This will mean
directly fitting observed light curves, as opposed to fitting folded
light curves (pulseshapes). Work in this direction is presently
underway, and we hope to present such fits in the near future.

In addition, aspects of the approach employed here may also be
relevant for other classes of pulsating stars. For instance, standard
models of Cepheids and RR~Lyrae stars indicate that the convective
turnover timescale in their outer \ion{H}{1} zones should be about
one-tenth that of their pulsation periods, so the assumptions made in
section~\ref{ass} concerning the instantaneous response of the
convection zone to the pulsations should still be valid, although
changes in the radius of the star also need to be taken into account.
In particular, we believe that our approach could possibly provide an
alternate explanation for the so-called ``phase lag'' seen in
Cepheids.  This phase lag is the observed difference in phase between
the time of maximum light and that of minimum radius. The standard
explanation due to \citet{Castor68} assumes that the hydrogen
ionization zone is completely radiative, whereas it is very likely
strongly convective.  Such an investigation could lead, for instance,
to independent constraints (e.g., mass, depth) on the hydrogen
convection zones in these stars.

Finally, the space mission MOST's \emph{non-detection} of solar-like
oscillations in the star Procyon \citep{Matthews04} calls into
question our detailed understanding of convection in stars even
slightly different than the Sun. This indicates that the present model
of solar convection and pulsation may not scale in the way which is
expected when we extrapolate from the solar case, and is just one of
many pieces of evidence that we still have a long way to go in
understanding the physics of convection in stars.

\acknowledgements

The author would like to thank the anonymous referee for helpful
comments, as well as D. Koester, D.~O. Gough, D.~E.  Winget, and T.~S.
Metcalfe for useful discussions, and F. Mullally, S.~J.
Kleinman, and S.~O.  Kepler for providing some of the data analyzed in
section~\ref{fits}.


\end{document}